\documentclass[journal]{IEEEtran}
\usepackage{cite}
\usepackage{amsmath,amssymb,amsfonts}
\usepackage{algorithmic}
\usepackage{textcomp}
\usepackage{soul}
\usepackage[colorlinks=true,linkcolor=blue, citecolor=blue]{hyperref}%
\usepackage{balance}
\usepackage{multicol}
\usepackage{multirow}
\usepackage{array}
\usepackage[table,x11names]{xcolor}
\newcolumntype{P}[1]{>{\centering\arraybackslash}p{#1}}
\newcolumntype{M}[1]{>{\centering\arraybackslash}m{#1}}
\usepackage{supertabular}
\usepackage{mathtools}
\usepackage{graphicx, caption}%, subcaption}
\usepackage{blkarray, bigstrut}
 \newcommand*\Laplace{\mathop{}\!\mathbin\bigtriangleup}
 
\usepackage{subfig}
\usepackage{stfloats}
\newlength\savedwidth

\newlength\savewidth

\def\BibTeX{{\rm B\kern-.05em{\sc i\kern-.025em b}\kern-.08em
		T\kern-.1667em\lower.7ex\hbox{E}\kern-.125emX}}
\begin{document}
%\title{Coordinated  Energy Management Framework for Profit Maximization in Active Power Networks\vspace{-0.5cm}}
\title{Optimally Coordinated Energy Management Framework for Profit
	Maximization Considering Dispatchable and Non-Dispatchable Energy Resources}
%\title{Coordinated Framework for  Optimal Energy Management of Active Distribution Networks}
%\title{Coordinated Energy Management Framework for Profit Maximization \vspace{-0.91cm}}
%\title{Coordinated Optimization Framework for Profit Maximization in Active Distribution Networks\vspace{-0.93cm}}
\author{Rayees~Ahmad~Thokar,~\IEEEmembership{Student~Member,~IEEE,}
	Nikhil~Gupta,~\IEEEmembership{Senior~Member,~IEEE,}
	K.~R.~Niazi,~\IEEEmembership{Senior~Member,~IEEE,}
	Anil~Swarnkar,~\IEEEmembership{Senior~Member,~IEEE,}
	Nand~K.~Meena,~\IEEEmembership{Member,~IEEE,}
	and~Jin~Yang,~\IEEEmembership{Senior~Member,~IEEE}% <-this % stops a space
\vspace{-0.8cm}}
%\thanks{Rayees Ahmad Tokar, Nikhil Gupta, Anil Swarnkar, and K. R. Niazi are with the Department of Electrical Engineering, Malaviya National Institute of Technology, Jaipur, 302017, India}% <-this % stops a space
%\thanks{Nand K. Meena is with School of Engineering and Applied Science, Aston University, Birmingham, B4 7ET, U.K.}% <-this % 
%\thanks{Jin Yang is with James Watt School of Engineering, University of Glasgow, G12 8LT, U.K.}% <-this % stops a space
%\vspace{-1cm}}
\markboth{IEEE Systems Journal \LaTeX\ Class Files,~Vol.~00, No.~00, April~2020}%
{Shell \MakeLowercase{\textit{et al.}}: Bare Demo of IEEEtran.cls for IEEE Journals}
\maketitle
\begin{abstract}
Contemporary distribution network can be seen with diverse dispatchable and non-dispatchable energy resources. The coordinated scheduling of these dispatchable resources with non-dispatchable resources can provide several techno-economic and social benefits. Since, battery energy storage systems (BESSs) and microturbine (MT) units are capital intensive, a thorough investigation of their coordinated scheduling on pure economic basis will be an interesting and challenging task while considering dynamic electricity price and uncertainty handling of non-dispatchable resources and load demand. This paper proposes a new methodology for optimal coordinated scheduling of BESSs and MT units considering existing renewable energy resources and dynamic electricity price to maximize daily profit function of the utility by employing a recently explored modified African buffalo optimization (MABO) algorithm. The key attributes of the proposed methodology are comprised of mean price-based adaptive scheduling embedded within a decision mechanism system (DMS) to maximize arbitrage benefits. DMS keeps a track of system states as a-priori thus guides the artificial intelligence based solution technique for sequential optimization. This may also reduce the computational burden of complex real-life engineering optimization problems. %As a result of this the computational burden of complex real-life engineering optimization problems can be reduced. 
Further, a novel concept of fictitious charges is proposed to restrict the counterproductive operational management of BESSs. The application results investigated and compared on a benchmark 33-bus test distribution system highlights the importance of the proposed methodology. 
\end{abstract}
	\begin{IEEEkeywords}
Adaptive scheduling, arbitrage benefits, battery energy storage, energy management, fictitious charges, nature-inspired optimization. \vspace{-0.3cm}  
		%e-mail to keywords@ieee.org or visit \underline
		%{http://www.ieee.org/organizations/pubs/ani\_prod/keywrd98.txt}
	\end{IEEEkeywords}
%	\titlepgskip=-15pt
%	\maketitle
%
\section*{Nomenclature}
	\subsection{parameters and sets}
	\renewcommand{\arraystretch}{0.95}
	\begin{supertabular}[l]{p{35pt} p{209pt}}
		$\overline{{{E}_{G}}}$ & Mean price of electricity (\$/kWh). \\
		${{E}_{\text{SPV}}}$ & Energy purchase price from SPVs (\$/kWh). \\
		${{E}_{\text{WT}}}$ & Energy purchase price from WTs (\$/kWh). \\
		${{E}_{\text{MT}}}$ & Fuel charges against MT power generation (\$/kWh).\\
		$E_{\text{BESS}}^{F}$ & Fictitious charges against BESS dispatch (\$/kWh). \\	
		$H$ 							& Round-trip efficiency of BESSs (\%). \\	
		$I_{j}^{\text{max}}$ & Ampacity of $j$th line (A). \\	
		${{M}_{\text{MT}}}$ & O\&M charges for MT (\$/kWh). \\
		${{M}_{\text{BESS}}}$ & O\&M charges for BESS (\$/kWh). \\			
		${{N}_{\text{s}}}$ & Total system states. \\
		${{N}_{\text{SPV}}}$ & Total number of SPV units. \\
		${{N}_{\text{WT}}}$ & Total number of WT units. \\
		${{N}_{\text{MT}}}$ & Total number of MT units. \\
		${{N}_{B}}$ & Total number of BESS units. \\
		$P_{\text{MT},r}^{R}$ & Rated power output of $r$th MT unit  (kW). \\	
		$P_{\text{BESS},b}^{\text{max},C}$ & Maximum charging limits of $b$th BESS (kW). \\
		$P_{\text{BESS},b}^{\text{max},D}$ & Maximum discharging limits of $b$th BESS (kW). \\		
		${\text{SOC}}_{I}$ & Initial specified State of Charge of BESSs (\%). \\
		${\text{SOC}}_{b}^{\text{min/max}}$ & Minimum/maximum SOC of $b$th BESS (\%). \\
		$T_{\text{BESS}}^{C}$ & Set of tentative hours for BESS charging mode. \\
		$T_{\text{BESS}}^{S}$ & Set of tentative hours for BESS standby mode. \\
		$T_{\text{BESS}}^{D}$ & Set of tentative hours for BESS discharging mode. \\
		$T_{\text{MT}}^{D}$ & Set of tentative hours for MT dispatch mode. \\		
		$T$ & Set of system states. \\
		$V_{{}}^{\text{max/min}}$ & Upper and lower bounds of node voltage (p.u.). \\					
		$W_{\text{BESS}\text{,}b}^{R}$ & Rated capacity of $b$th BESS (kWh). \\	
		${{\Omega }_{N}}$ & Set of system nodes. \\
		${{\Omega }_{E}}$ & Set of system lines. \\
		${{\Omega }_{\text{SPV}}}$ & Set of SPV units. \\
		${{\Omega }_{\text{WT}}}$ & Set of WT units. \\
		${{\Omega }_{\text{MT}}}$ & Set of MT units. \\
		${{\Omega }_{B}}$ & Set of BESS units. \\
		${{\eta }_{C/D}}$ & Charging/discharging efficiency of BESSs (\%). \\[-0.4cm]	
		%$\Re $ & Real part of a complex number. \\[-0.4cm]	
	\end{supertabular}
    \subsection{variables}
   \renewcommand{\arraystretch}{0.95}
	\begin{supertabular}[l]{p{55pt} p{190pt}}
	 ${{E}_{G}}\left( \Delta {{t}_{i}} \right)$ & Grid energy price during $i$th state ($\$$/kWh). \\
	 ${{E}_{C}}\left( \Delta {{t}_{i}} \right)$ & Customers energy price at $i$th state (\$/kWh). \\
	 ${{I}_{j}}(\Laplace {{t}_{i}})$ & Current  in $j$th line during $i$th state (A). \\
	 $I_{1}^{{}}/{{V}_{1}}(\Laplace {{t}_{i}})$ & Secondary current (A)/voltage (V) magnitudes of grid sub-station transformer at $i$th state. \\
	 ${{P}_{\text{SPV},p}}(\Laplace {{t}_{i}})$ & Output of $p$th SPV unit during $i$th state (kW). \\
	 ${{P}_{\text{WT},q}}(\Laplace {{t}_{i}})$ & Output of $q$th WT unit during $i$th state (kW). \\
	 ${{P}_{\text{MT},r}}(\Laplace {{t}_{i}})$ & Output of $r$th MT unit during $i$th state (kW). \\
	 $P_{\text{BESS},b}^{C/D}(\Laplace {{t}_{i}})$ & Charging/discharging dispatch of $b$th BESS during $i$th state (kW). \\
	 ${{P}_{\text{loss}}}(\Laplace {{t}_{i}})$ & Feeder power losses during $i$th state (kW). \\
	 ${{P}_{D}}(\Laplace {{t}_{i}})$ & Customers load demand during $i$th state (kW). \\
	 ${{P}_{\text{rev}}}(\Laplace {{t}_{i}})$ & Back-feed to grid substation in $i$th state (kW). \\
	 ${{P}_{G}}(\Laplace {{t}_{i}})$ & Grid power generation during $i$th state (kW) . \\
	 ${\text{SOC}}_{b}(\Laplace {{t}_{i}})$ & SOC of $b$th BESS during $i$th state (\%). \\
	 ${{V}_{k}}(\Laplace {{t}_{i}})$ & Voltage at $k$th node during $i$th state (p.u.). \\
	 $\Delta {{t}_{i}}$ & Duration of $i$th system state . \\
	 ${{\delta }_{1}/{\delta }_{2}}(\Laplace{{t}_{i}})$ & Voltage angle at bus 1\&2 during $i$th state. \\%[-0.3cm]	
    \end{supertabular}
%	\subsection{Parameters}
%	\begin{supertabular}[l]{p{30pt} p{195pt}}
%	\end{supertabular}
	%
	\section{Introduction}\label{sec1}	
	Modern distribution systems can be seen with high penetration of distributed generations (DGs) on account of several techno-economic and social concerns. These DG units may exist in a combination of non-dispatchable sources, i.e. solar photovoltaics (SPVs), wind turbines (WTs), etc. and dispatchable sources such as micro turbines (MTs), battery energy storage systems (BEESs), etc. The simultaneous optimal scheduling of deployed BESSs and MT units in the presence of uncertain renewable energy sources (RESs) helps in exploiting techno-economic benefits for utilities and other stake holders (DG owners and consumers) leading to  deferral of assets upgrade, reliability enhancement, improvement in power quality, price and energy arbitrage benefits, cost minimization etc. \cite{1300713, 7973037, 7879188}. However, the escalating presence of BESSs requires scheduling and optimizing their coordination and management \cite{6652316,6695368}. This is because such coordinated management can absorb intermittency of renewable power generation besides significant arbitrage benefits under dynamic electricity pricing owing to their flexibility of dispatching energy at appropriate time. Since BESSs are capital intensive, their coordinated management must involve the optimization of daily profit function of the utility. Therefore, arbitrage benefits and desired operating strategy of dispatchable sources are the key issues that must be taken care of while solving the decision problem for profit maximization. Nevertheless, the problem is challenging on account of dynamic behavior of the generation from RESs, demand, and electricity market that lead to a complex process which needs adaptive strategies to deal with BESSs \cite{reddy2011linden}. The consideration of dynamic electricity price may further add computational burden to the optimization technique on account of dynamically varying system states with large number of state variables \cite{Hannah:2011:ADP:3104482.3104525}.
\par	Several works \cite{7339477,1519727,5438853,6797967,6310769,7321810,7422915} have been reported for coordinated management of energy storage devices with diverse DERs using different charging/discharging strategies and artificial intelligence (AI) based optimization techniques. In most of the literature, the scheduling of storage technologies is either governed by the objective function to be optimized or sometimes by time dependent fixed operating strategies. Some of the research work are as follows. A Markov decision processes (MDP) based method \cite{7339477} is employed to optimally schedule energy storage devices in distribution systems with renewable integration to minimize system losses and overall energy cost. 
%In \cite{1425561}, the optimal operation strategies and capacities of the multiple BESS technologies are determined for sequentially evaluating the impact of the specific costs of storage technologies on the overall net present value (NPV). 
The scheduling of diverse power system energy storages is proposed in \cite{1519727} using deterministic and stochastic approaches. Non-dominated sorting genetic algorithm is used as an optimization technique.  In \cite{5438853}, optimal management and allocation of BESSs in distribution systems is proposed to mitigate wind curtailment, loss minimization and feasibility enhancement. However, the charging and discharging dispatch of BESSs is fixed and is governed by off-peak and on-peak hours respectively. A trade-off between energy losses and the utilization of available energy reserves is presented in \cite{6797967} by proposing a dynamic offset policy (DO) for optimal scheduling of energy storage. It has been shown that, DO outperform existing heuristics of \cite{6310769} and can also be adapted for maximization of profit in a dynamic pricing environment. In \cite{7321810}, a strategy for optimal integration of BESSs is proposed while considering loss reduction and voltage regulation as objectives. The capacity adjustment technique of BESSs is proposed in \cite{7422915} by considering optimal accommodation and scheduling of BESSs while minimizing system losses and NPV of BESSs and distribution network. In \cite{Hannah:2011:ADP:3104482.3104525,7973220,7741131}, dynamic programming is employed for optimal scheduling of BESSs under uncertain environment. The problem is evaluated for minimization of cost function to ensure flexible and economic utilization of BESSs. The authors concluded that practically large scale BESS utilization can be achieved while optimizing consumption from RESs.
\par The optimal utilization of BESSs involves charging/discharging intervals in accordance to dynamic demand, renewable power generation, and pricing signal. In most of the above mentioned works, these intervals are kept fixed on the basis of aggregated demand curve. Such battery operation strategy may lead to low profit and increased energy losses, since the profiles of renewable penetration, demand and electricity price vary on daily basis \cite{6387346}. However, a flexible and adaptive operational strategy for BESSs together with uncertainty modelling of load demand and RES generation may provide better utilization of existing energy resources and enhanced arbitrage benefits.
\par Energy storages are capital intensive and their useful life is usually less than the other DERs. The scheduling of BESS in dynamic electricity pricing environment therefore essentially considers arbitrage benefits, which have not been fully exploited, in order to justify their installation.  The arbitrage benefits have been investigated by limited researchers \cite{6345242,en9010012,8353004,JANNESAR2018957,en12071231}. In Ref. \cite{6345242}, a simple price thresholds policy is utilized to determine the optimal operation control of the storage technologies for assessing the arbitrage benefits. However, the solution is independent of the state of charge (SOC) and also empties storage system completely at the end of operation which may seriously affect their longevity. A charge/discharge scheduling method to evaluate the arbitrage potential of the BESS systems is suggested in \cite{en9010012} to maximize the benefits of BESS owners. The study introduces dynamic pricing profile instead of load profile to determine dispatch hours of storages while considering several assumptions. In \cite{8353004}, a two stage stochastic framework is formulated for optimal allocation and scheduling of BESS and MT in a microgrid while minimizing the cost function through BESS arbitrage benefits and enhancing the reliability and robustness of the system. The authors' concluded that by utilizing large size MT economically more feasible solutions will be obtained. Recently, an economic model for simultaneous allocation and scheduling problem of BESS is proposed in \cite{JANNESAR2018957} while considering techno-economic and social objectives. The study highlights that a positive impact on the high penetration of RESs, energy arbitrage, power losses, node voltage profile and emission can be achieved using proposed methodology. Ref. \cite{en12071231} proposed day-ahead operational planning for optimal scheduling of BESSs using multi-period optimal power flow considering intermittency of non-dispatchable DGs.
\par From the aforesaid discussion, it can be summarized that the scheduling problem of BESSs with MT unit must be coordinated with the existing RESs considering dynamic pricing signal. Moreover, BESSs are considered either utility or third party owned. However, when the BESSs are utility owned, the daily profit function has not been taken into consideration. If BESSs are utility owned, optimizing utility’s daily profit function may be of prime concern as they are capital intensive.  Such coordinated scheduling can be managed in such a way to achieve several tangible benefits, i.e. optimum utilization of energy resources, price and energy arbitrage, power loss reduction, peak-load shaving, load profile improvement, etc. However, the task is highly challenging on account of coordination and control of the charging, standby and discharging status of BESSs under dynamically varying signals of load demand, power generation and price over the numerous system states. The profit function is discontinuous, non-linear and encompasses several system states together. Therefore, it can’t be efficiently optimized with less computational burden using an AI based nature-inspired meta-heuristic technique, like GA or PSO, unless is supplemented by suitable means to restrict enormous problem search space. A flexible, adaptive and suitably tailored methodology along with knowledge base driven AI based technique may provide compromising solutions. 
\par This paper proposes a new methodology for optimal coordinated management of utility owned BESSs with MT unit under dynamic electricity pricing environment to maximize daily profit function (DPF) of the utility while fully exploiting existing DERs in the distribution system. The DPF is optimized using recently explored swarm intelligence-based optimization technique namely modified African buffalo optimization (MABO) with high investigation and utilization competencies while considering intermittency and variability in renewable power generation and load demand. The salient contributions of the proposed methodology are the mean price-based adaptive scheduling (MPAS) of BESSs, decision mechanism system (DMS) and fictitious charges (FCs). MPAS provides price-based dynamic charging and discharging intervals for BESSs. DMS tracks all the system states to gather and process the information as a priori, thus guides MABO for sequential optimization. This may reduce computational burden and may also enhance efficacy of the AI based solution technique. In addition, the methodology employs a novel concept of fictitious charges (FCs) to check uneconomic operational management of BESSs. All these features facilitate desired operation of utility owned BESSs while optimizing profit function. These features have not been yet explored to the best of authors' knowledge. The analysis and comparison of the application results on a benchmark 33-bus test distribution system emphasize the significance of the proposed methodology. 
%	
%	The structure of the continuing paper is planned as follows. Section \ref{proposed mean price-based} presents the proposed scheduling methodology. The section also addresses in depth decision mechanism system and its associated parts, followed by problem formulation in Section \ref{section:problem formulation}. Solution methodology for coordinated scheduling is discussed in Section \ref{section:solution methodology}. Numerical results and discussion are presented in Section \ref{section:simulation results and discussion} to substantiate the effectiveness of the proposed scheduling. Finally, the concluding remarks are drawn in Section \ref{section:conclusions}.
\section{Proposed mean price-based adaptive scheduling}\label{proposed mean price-based}
\subsection{Proposed Energy Management Framework}\label{section:EMF}
Modern day distribution systems are having integrated dispatchable and non-dispatchable distributed energy resources, which may include SPVs, WTs, BESSs and MTs, etc. The integration of diverse DER technologies is vital for the optimal operation of contemporary distribution systems. However, the proper coordination among multiple DERs is very important from both the operational and economic efficiency point of view of the contemporary distribution systems. For better economic efficiency, RESs must be fully tapped and utilized locally \cite{8353004,6662462}. In this context, the optimal management of energy storage system (ESS) has been acknowledged as an amicable solution. Now-a-days, BESSs have received more and more attention and application due to technical and economic advantages \cite{6305498}. The advantages associated on account of its ability to deliver bi-directional power in the distribution network. This ability is crucial while considering dynamic electricity price as BESSs can store energy during off-peak hours and deliver the same during peak hours. However, if MT is used in coordination with BESSs better results may be achieved while considering forecasting errors in load demand and power generation from RESs \cite{8353004}. With these concerns, BESSs and MT unit can be optimally scheduled to maximize the techno-economic efficiency of the distribution systems while maintaining several operational constraints.
\par In this work, it has been assumed that SPVs and WTs are owned by the third party, i.e., distributed generation owners (DGOs), whereas MT unit and BESSs are utility owned. The utility therefore purchases energy from the grid under day-ahead dynamic electricity pricing environment, whereas from DGOs at the contract price. The utility in turn generates revenue by selling energy to the customers at a dynamic day-ahead price while preserving a definite fixed profit of margin. The MT unit can inject energy into the system during peak pricing hours, or otherwise, while keeping sufficient reserve, as in \cite{8353004}. 
\par The optimal scheduling of BESS with MT imposes challenges on account of intermittency of RESs, stochastic nature of load demand, charging/discharging status of BESS, forecasting errors in load demand and renewable power generation, etc. These issues essentially requires the consideration of dynamically varying system states which may be very large in number over the time frame of the problem, say over a sample day. The optimal scheduling of these components needs coordinated operation with RESs as well as with dynamic electricity pricing signal, if maximizing DPF. This imposes real challenges owing to optimal utilization of existing DERs. Such complex non-linear optimization problems can be effectively handled using AI-based nature-inspired meta-heuristic techniques but are highly computationally demanding due to simultaneous optimization of large number of system states, otherwise accuracy may suffer. Alternatively, sequential optimization may be adopted. However, sequential optimization evolves difficulties because the AI-based solution technique is optimizing the objective function for the current state while rest of the states remained unsighted. It may eventually results in loss of coordination control over the charging/discharging status of BESS while considering dynamic electricity pricing. It happened because the status of BESS will be decided by the profit function of that particular state without considering profit function of the day as the BESS is utility owned. This is highly undesirable as it may cause underutilization of BESS or alleviate battery lifetime. Thus sequential optimization essentially needs the support of a suitable decision based system to look after all system states and provide necessary inputs to the AI-based solution technique.
\subsubsection{Decision Mechanism System }\label{section:decision mechanism}
Several authors \cite{1300713,7973037,7879188,6652316, 6695368,7339477,7973220,6662462,6305498} proposed operating strategies for different DER technologies considering time as the decision vector to determine their optimal scheduling while considering dynamic nature of electricity pricing. However, price seems to be more direct, flexible and attractive decision vector alternative while evaluating the scheduling of DERs to maximize profit function. A decision mechanism system (DMS) is developed that suggests suitable heuristic-based charging/discharging strategy of BESS over the scheduling period, as a priori, while considering dynamic electricity price. In this mechanism, the mean price plays as the decision vector in deciding the charging/discharging status of BESS thus helps to avail price and energy arbitrage benefits. With this price decision vector, the possible power transactions from controlled DERs now can be assessed easily throughout the scheduled period. In this way, DMS guides the AI based solution technique for other system states. For instance, the decision about the charging and discharging mode of BESSs may be governed by the following respective equations:\vspace{-0.2cm}
	\begin{equation}
		\scalebox{0.9}[0.9]{$
	{{E}_{G}}\left( \Laplace {{t}_{i}} \right)<\overline{{{E}_{G}}}\,;\,\forall \Laplace {{t}_{i}}\in T_{\text{BESS}}^{C}
	$}
	\label{eqn1}
	\end{equation}
	\begin{equation}
	\scalebox{0.9}[0.9]{$
	{{E}_{G}}\left( \Laplace {{t}_{i}} \right)>\overline{{{E}_{G}}}\,;\,\forall \Laplace {{t}_{i}}\in T_{\text{BESS}}^{D}
	$}
	\label{eqn2}
	\end{equation}
\subsubsection{Algorithm for BESS management}\label{Subsection:algorithm for bESS}
The above mentioned equations merely decide the charging or discharging status of BESS. For economic operation, the scheduling of BESS over the sub-periods $T_{\text{BESS}}^{D}$ and $T_{\text{BESS}}^{C}$ should consider dynamic price to enhance DPF. This needs identification of those particular states where either charging or discharging of BESS is optimum. Furthermore, there is a need to address boundary conditions of BESS dispatches for theses selected states so that it can be supplied whenever demanded by the solution technique. Therefore, following simplified algorithm is suggested:
\begin{enumerate}
		\item Arrange all system states above the mean price in descending order of price, thus obtaining the priority of discharging states.
		\item Subsequently assign dispatch limit  $P_{\text{BESS},b}^{\text{max},D}$ for $b$th BESS, on priority, for these discharging states till SOC exhausted. However, the dispatch allocation for the final discharging state being selected may not be $P_{\text{BESS},b}^{\text{max},D}$, which can be evaluated using \eqref{eqn3}, as given bellow:
		 \begin{equation}
		 	\scalebox{0.9}[0.9]{$
		 \begin{split}
		 & \left( \frac{\left( \text{SOC}_{b}^{\text{max}}-\text{SOC}_{b}^{\text{min}} \right)W_{\text{BESS},b}^{R}-zP_{\text{BESS},b}^{\text{max},D}}{{{\eta }_{D}}} \right); \\ & z=\text{floor}\left( \frac{\left( \text{SOC}_{b}^{\text{max}}-\text{SOC}_{b}^{\text{min}} \right)W_{\text{BESS},b}^{R}}{P_{\text{BESS},b}^{\text{max},D}} \right)
		 \end{split}
		 $}
		 \label{eqn3}
		 \end{equation}
\item The dispatch limit for the remaining discharging states is set to zero.
\end{enumerate} 
While considering decisions about the charging of BESS, the same algorithm can be modified as described below:
 \begin{enumerate}
	 	\item Arrange all system states below the mean price in ascending order of price, thus obtaining the priority of charging states.
	 	\item Subsequently assign dispatch limit of charging $P_{\text{BESS},b}^{\text{max},C}$ for $b$th BESS, on priority, for these charging states till the upper limit of SOC reached. However, the dispatch allocation for the final charging state being selected may not be $P_{\text{BESS},b}^{\text{max},C}$, which can be evaluated using \eqref{eqn4}, as presented below:
	 	\begin{equation}
	 		\scalebox{0.9}[0.9]{$
	 	\begin{split}
	 	& \left( \left( \text{SOC}_{b}^{\text{max}}-\text{SOC}_{b}^{\text{min}} \right)W_{\text{BESS},b}^{R}-zP_{\text{BESS},b}^{\text{max},C} \right){{\eta }_{C}}; \\ & z=\text{floor}\left( \frac{\left( \text{SOC}_{b}^{\text{max}}-\text{SOC}_{b}^{\text{min}} \right)W_{\text{BESS},b}^{R}}{P_{\text{BESS},b}^{\text{max},C}} \right)
	 	\end{split}
	 	$}
	 	\label{eqn4}
	 	\end{equation}
	 	\item The charging limit for the remaining states is set to zero. 	
	\end{enumerate}
	In this way, DMS helps the optimization techniques to ensure full discharging/charging of BESS during highest price band/lowest price band of the day. System states where no power transaction of BESSs take place will be representing the standby mode. Such standby mode between the consecutive charging and discharging modes of BESS is very useful. It helps to alleviate the inconvenience caused to the operation of converters interconnected to the BESSs due to direct switching from charging mode to discharging mode and vice-versa \cite{7973220}. 
\subsubsection{Fictitious Charges (FCs)}
	Since the BESS is assumed to be utility owned, practically no revenue is received by the utility against BESS charging and no charges shall be paid by the utility while it delivers energy into the system. With this concern, BESS would not charge at all because then definite charges have to be paid by the utility to either grid or DGOs and that reduces profit function of that state. Moreover, BESS will be ready every time to release energy into the system as this reduces the charges to be paid by the utility against energy purchase from the grid or DGOs, thus increases profit function of that state. This certainly hampers the BESS management process. These difficulties have been overcome in proposed methodology by introducing a novel concept of fictitious charges (FCs). In fact, FCs will provide a watershed in optimization process while scheduling the BESSs for profit function maximization. FCs are suggested to impose on BESSs whenever transacting energy with the distribution system. With this ideology, the utility virtually receives revenue@FC against each unit charging of BESS. Similarly, the utility shall virtually pay@FC for each unit of energy being discharged from the BESS. The FC, $E_{\text{BESS}}^{F}$, is kept fixed at the mean price with the consideration of \eqref{eqn1} and \eqref{eqn2}. The action of introducing FCs avoids unwanted BESS scheduling as explained in Table \ref{Table:1}.
\begin{table}
\caption{Action of fictitious charges on BESS scheduling}
\label{Table:1}
\centering
	\renewcommand{\arraystretch}{1.1}
\begin{tabular}{|p{1.05cm}|p{3.3cm}|p{3.2cm}|}
\hline
\rowcolor{gray!30} Price Relation & Charging of BESS & Discharging of BESS\\ 
\hline
\rowcolor{lime!30} $E_{\text{BESS}}^F> E_{G}(\Laplace t_{i});$ $\forall \Laplace t_i\in T_{\text{BESS}}^{C}$ & Facilitates as the utility receives more amount from BESS against the energy supplied for their charging than what it pays to purchase the same. & Restrains as the utility receives less revenue from the customers than what it has to pay to BESSs for supplying this energy into the system. \\
\hline
\rowcolor{orange!30} $E_{\text{BESS}}^F<E_{G}(\Laplace t_{i});$ $\forall \Laplace t_i\in T_{\text{BESS}}^{D}$ & Restrains as the utility receives less revenue from the customers than what it has to pay to BESSs for supplying this energy into the system.    & Facilitates as the utility receives more amount from BESS against the energy supplied for their charging than what it pays to purchase the same.\\ 
\hline
		\end{tabular}
	\vspace{-0.6cm}
	\end{table}
	FCs are imposed simply to prevent undesirable operation of BESS during optimization, however, these are deducted before evaluating fitness of the objective function.	
\subsubsection{Operational Strategy for MT}
The above mentioned algorithm can be used to determine the possible optimal dispatch of MT unit when dynamic prices are higher than its cost of power generation. But, the purpose of integrating MT unit in the system is different than that of integrating BESSs. The main purpose behind the integration of MT unit is to absorb the power imbalance between contract demand and local power generation on account of forecasting error in the power generation from RESs and load demand \cite{8353004}. However, the integration of MT is capital intensive therefore economic aspects should be duly considered for better utilization, besides power balance. Nevertheless, sufficient reserve should be kept to match forecasting errors \cite{7973037}. As MT is a controlled DG, the decisions about the reserve requirement and scheduling strategy will be on the sole discretion of utility. For the given decisions, the possible dispatches and corresponding dispatch period for MT scheduling while optimizing DPF can be determined using the algorithm suggested in Subsection \ref{Subsection:algorithm for bESS}. This information is also transferred to DMS to guide the AI-based solution technique. 
\subsection{Renewable Generation and Load Consumption Modelling }\label{section:RELC Modeling}
As discussed, the intermittency and variability in renewable power generation and load demand imposes certain challenges on performance of distribution networks. Therefore, in such a volatile environment, for significant improvement in operation performance of DER integrated distribution networks and to achieve more realistic results, an effective way of handling and modelling uncertain data becomes a primary concern for the network operator.
\par Numerous literary works are presented to handle the intermittent and variable nature of renewable generation and load demand through several stochastic approaches \cite{atwa2011probabilistic,malekpour2012multi,zio2015monte,jadoun2018integration}. However, these approaches are computationally intense and intractable due to the estimation of the probability distribution of uncertain data and their dependence on sampled scenarios of the uncertainty realizations\cite{wang2014robust,8272222,gorissen2015practical}. In \cite{wang2014robust}, authors’ introduced a simple deterministic approach (polyhedral uncertainty sets) to tackle the intermittency of RES generation and load consumption more competently than the conventional stochastic approaches. The approach requires fewer information of the uncertainty set like the mean, lower and upper bounds of the uncertain data which are easily obtainable from the historical data and thus provides computational tractability for the uncertainty sets. This approach of intermittency handling suffers from conservativeness of optimal solution which is mitigated in \cite{8272222} by introducing self-adaptive polyhedral uncertainty sets.
\par In \cite{8272222}, the synthetic data is generated from available annual data by utilizing data spread (DS) and budget of uncertainty (BOU). The DS and BOU are made dynamic unlike \cite{wang2014robust} where they are taken as fixed while generating uncertainty sets. These are constructed by using hourly mean and standard deviation (SD) of the monthly data. For detailed information the reader may refer \cite{8272222}. However, while constructing BOU there arises a probability of over-constraint or under-constraint problem for the months with low solar/wind generation and less time varying load demand or for the months with high solar/wind generation and high time varying load demand respectively. And consequently accuracy of the synthetic data generated may be lost. In the present work, the authors’ have defined BOU as the mean and SD of the monthly data on daily basis in addition to hourly basis. This small modification in BOU may successfully mitigate over-constraint or under-constraint problem and thereby enhancing accuracy of the synthetic data. Further, the modified modelling approach considers in hand availability of annual data for solar and wind power generation and load demand \cite{nrel,posoco}.
\par The mathematical expression of the modified uncertainty set  $W_{y,m}^{SPV}(\vartriangle{{t}_{i}})$ for SPV power generation can be defined as
\begin{equation}
\scalebox{0.89}[0.89]{$
\begin{split}
 W_{y,m}^{SPV}\left(\vartriangle{{t}_{i}} \right)= \chi _{n,y,m}^{SPV}\left( \vartriangle {{t}_{i}} \right)\in {{R}^{SPV}}:\underline{\omega }_{n,y,m}^{SPV}\left( \vartriangle {{t}_{i}} \right) \\\le \chi _{n,y,m}^{SPV}\left(\vartriangle {{t}_{i}} \right)\le \overline{\omega }_{n,y,m}^{SPV}\left( \vartriangle {{t}_{i}} \right); 
 \forall \ n\in {{\Omega }_{N}},\forall \vartriangle {{t}_{i}}\in T
\end{split}
$}
\label{eqn5a}
\end{equation}
where,
\begin{equation}
\scalebox{0.9}[0.9]{$
	\begin{split}
\underline{\omega }_{n,y,m}^{SPV}\left( \vartriangle {{t}_{i}} \right)=\omega _{n,y,m}^{SPV}\left( \vartriangle {{t}_{i}} \right)-k\sigma _{n,y,m}^{SPV}\left( \vartriangle {{t}_{i}} \right),\,\\ \overline{\omega }_{n,y,m}^{SPV}\left( \vartriangle {{t}_{i}} \right)=\omega _{n,y,m}^{SPV}\left( \vartriangle {{t}_{i}} \right)+k\sigma _{n,y,m}^{SPV}\left( \vartriangle {{t}_{i}} \right) \\ 
\, \\ 	
\end{split}
$}
\label{eqn6a}
\end{equation}
such that,
\begin{equation}
\scalebox{0.9}[0.9]{$
	\begin{split}
\text{ }\underline{\mu }_{n,y,m}^{SPV}\left( \vartriangle {{t}_{i}} \right)\,\le \,\hat{\chi }_{n,y,m}^{SPV}\left( \vartriangle {{t}_{i}} \right)\le \overline{\mu }_{n,y,m}^{SPV}\left( \vartriangle {{t}_{i}} \right); \\ \underline{\mu }_{n,y,m}^{SPV}\left( \vartriangle {{t}_{i}} \right)=\mu _{n,y,m}^{SPV}\left( \vartriangle {{t}_{i}} \right)-k\hat{\sigma }_{n,y,m}^{SPV}\left( \vartriangle {{t}_{i}} \right)\,,\,\\ \overline{\mu }_{n,y,m}^{SPV}\left( \vartriangle {{t}_{i}} \right)=\mu _{n,y,m}^{SPV}\left( \vartriangle {{t}_{i}} \right)+k\hat{\sigma }_{n,y,m}^{SPV}\left( \vartriangle {{t}_{i}} \right) \\
\end{split}
$}
\label{eqn7a} 
\end{equation}
In \eqref{eqn5a}--\eqref{eqn7a}, the available annual data and data to be synthesized are represented by $\omega$-terms and $\chi$-terms respectively. The DS and BOU are described by the intervals [$\underline{\omega }_{n,y,m}^{SPV}(\vartriangle{{t}_{i}})$, $\overline{\omega }_{n,y,m}^{SPV}(\vartriangle{{t}_{i}})$] and [$\underline{\mu }_{n,y,m}^{SPV}(\vartriangle {{t}_{i}})$, $\overline{\mu }_{n,y,m}^{SPV}(\vartriangle {{t}_{i}})$] at node $n$ during $i$th system state for $m$th month of the year as shown in \eqref{eqn5a} and \eqref{eqn6a} respectively. Further, the value of $k$ (user-defined coefficient) is taken as unity, as in \cite{8272222}. Likewise, the uncertainty sets for WT power generation and load demand can also be generated from the available data. For further details about the basic steps involved in generating uncertainty sets Ref. \cite{wang2014robust,8272222,gorissen2015practical} may be referred.	
\section{Problem Formulation}\label{section:problem formulation}
The objective function is to maximize the profit function for optimal scheduling of BESSs and MT unit while considering $i$th system state, which is formulated as:
    \begin{equation}
        \scalebox{0.9}[0.9]{$
    \begin{split}
    \text{Max}\,\text{OF}(\Laplace {{t}_{i}})=R(\Laplace {{t}_{i}})-P(\Laplace {{t}_{i}})-\text{FC}(\Laplace {{t}_{i}})
    \end{split}
    $}
    \label{eqn5}
    \end{equation} 
    where,
    \begin{equation}
    \scalebox{0.9}[0.9]{$
    \begin{split}
 R\left( \Laplace {{t}_{i}} \right)=\,{{P}_{D}}(\Laplace {{t}_{i}}){{E}_{C}}(\Laplace {{t}_{i}})\Laplace {{t}_{i}}+{{P}_{\text{loss}}}(\Laplace {{t}_{i}}){{E}_{G}}(\Laplace {{t}_{i}})\Laplace {{t}_{i}}\\ +E_{\text{BESS}}^{F}\sum\limits_{b=1}^{{{N}_{B}}}{P_{\text{BESS},b}^{C}(\Laplace {{t}_{i}})\Laplace {{t}_{i}}}
    \end{split}
    $}
    \label{eqn6}
    \end{equation}
    \begin{equation}
%    \small
    \scalebox{0.88}[0.89]{$
    \begin{split}
 P\left( \Laplace {{t}_{i}} \right)=\,{{P}_{G}}(\Laplace {{t}_{i}}){{E}_{G}}(\Laplace {{t}_{i}})\Laplace {{t}_{i}}+  {{E}_{\text{SPV}}} \sum\limits_{p=1}^{{{N}_{\text{SPV}}}}{{{P}_{\text{SPV},p}}(\Laplace {{t}_{i}})\Laplace {{t}_{i}}}\\ 
 + {{E}_{\text{WT}}}\sum\limits_{q=1}^{{{N}_{\text{WT}}}}{{{P}_{\text{WT},q}}(\Laplace {{t}_{i}})\Laplace {{t}_{i}}}+   {{E}_{\text{MT}}}\sum\limits_{r=1}^{{{N}_{\text{MT}}}}{{{P}_{\text{MT},r}}\left( \Laplace {{t}_{i}} \right)\Laplace {{t}_{i}}}\\
  +  {{M}_{\text{MT}}}\sum\limits_{r=1}^{{{N}_{\text{MT}}}}  {{{P}_{\text{MT},r}}(\Laplace {{t}_{i}})\Laplace {{t}_{i}}}+  {{M}_{\text{BESS}}} \sum\limits_{b=1}^{{{N}_{B}}}\Big(P_{\text{BESS},b}^{C}(\Laplace {{t}_{i}}) \\+  P_{\text{BESS},b}^{D}(\Laplace {{t}_{i}}) \Big) \Laplace {{t}_{i}}+  E_{\text{BESS}}^{F}\sum\limits_{b=1}^{{{N}_{B}}}{P_{\text{BESS},b}^{D}(\Laplace {{t}_{i}})\Laplace {{t}_{i}}} \\ 
     \end{split}
     $}
    \label{eqn7}
    \end{equation}
    \begin{equation}
        \scalebox{0.9}[0.9]{$
    \begin{split}
 \text{FC}\left( \Laplace {{t}_{i}} \right)= E_{\text{BESS}}^{F}\sum\limits_{b=1}^{{{N}_{B}}}{P_{\text{BESS},b}^{C}(\Laplace {{t}_{i}})\Laplace {{t}_{i}}}\\
 -E_{\text{BESS}}^{F} \sum\limits_{b=1}^{{{N}_{B}}}{P_{\text{BESS},b}^{D}(\Laplace {{t}_{i}})\Laplace {{t}_{i}}} 
    \end{split}
    $}
    \label{eqn8}
    \end{equation}
    Equation \eqref{eqn6} denotes the revenue of utility against the sale of energy, \eqref{eqn7} shows the payments by the utility against purchasing power from the grid and DGOs and \eqref{eqn8} represents FCs imposed on BESS energy transactions during ith state. The objective function defined by \eqref{eqn5} is optimized for each state and then the DPF is evaluated as below: 
    \begin{equation}
        \scalebox{0.9}[0.9]{$
    \text{O{F}}_{1}=\sum\limits_{i=1}^{{{N}_{S}}}{\text{OF}\left( \Laplace {{t}_{i}} \right)}
    $}
    \label{eqn9}
    \end{equation}
    Subjected to the following equality and inequality constraints 
    \begin{equation}
\scalebox{0.9}[0.9]{$
    \begin{split}
{{P}_{G}}(\Laplace {{t}_{i}})+\sum\limits_{p=1}^{{{N}_{\text{SPV}}}}{{{P}_{\text{SPV},p}}(\Laplace {{t}_{i}})}+\sum\limits_{q=1}^{{{N}_{\text{WT}}}}{{{P}_{\text{WT},q}}(\Laplace {{t}_{i}})}
+\sum\limits_{r=1}^{{{N}_{\text{MT}}}}{{{P}_{\text{MT},r}}(\Laplace {{t}_{i}})}\\
\mp \sum\limits_{b=1}^{{{N}_{B}}}{P_{{\text{BESS}},b}^{C/D}(\Laplace {{t}_{i}})}
-{{P}_{D}}(\Laplace {{t}_{i}})-{{P}_{\text{loss}}}(\Laplace {{t}_{i}}) =0;
~~\forall \Laplace {{t}_{i}}\in T_{\text{BESS}}^{C/D},\\ 
\forall \,p\in {{\Omega }_{\text{SPV}}},\forall \,q\in {{\Omega}_{\text{WT}}},\forall \,r\in {{\Omega }_{\text{MT}}},\forall \,b\in {{\Omega }_{B}}
    \end{split}
    $}
    \label{eqn10}
    \end{equation}
%    \begin{equation}
% \scalebox{0.9}[0.9]{$    
%    \begin{split}
%{{P}_{G}}(\Laplace {{t}_{i}})+\sum\limits_{p=1}^{{{N}_{\text{SPV}}}}{{{P}_{\text{SPV},p}}(\Laplace {{t}_{i}})}+\sum\limits_{q=1}^{{{N}_{\text{WT}}}}{{{P}_{\text{WT},q}}(\Laplace {{t}_{i}})}+\sum\limits_{r=1}^{{{N}_{\text{MT}}}}{{{P}_{\text{MT},r}}(\Laplace {{t}_{i}})}\\
%+\sum\limits_{m=1}^{{{N}_{B}}}{P_{\text{BESS},m}^{D}(\Laplace {{t}_{i}})} -{{P}_{D}}(\Laplace {{t}_{i}})-{{P}_{\text{loss}}}(\Laplace {{t}_{i}}) =0;\\
% \forall \Laplace {{t}_{i}}\in T_{\text{BESS}}^{D},\forall \,p\in {{\Omega }_{\text{SPV}}},\forall \,q\in {{\Omega }_{\text{WT}}},\forall \,r\in {{\Omega }_{\text{MT}}}, \forall \,m\in {{\Omega }_{B}}
%    \end{split}
%    $}
%\label{eqn11}
%    \end{equation}
    \begin{equation}
     \scalebox{0.9}[0.9]{$    
    \begin{split}
    \left| V_{{}}^{\min } \right|\le \left| {{V}_{k}}(\Laplace {{t}_{i}}) \right|\le \left| V_{{}}^{\max } \right|;\,\forall \Laplace {{t}_{i}}\in T,\forall k\in {{\Omega }_{N}}
    \end{split}
    $}
    \label{eqn12}
    \end{equation}
    \begin{equation}
         \scalebox{0.9}[0.9]{$    
    {{I}_{j}}(\Laplace {{t}_{i}})\le I_{j}^{\max};\,\forall \Laplace {{t}_{i}}\in T,\forall j\in {{\Omega }_{E}}
    $}
    \label{eqn13}
    \end{equation}
    \begin{equation}
            \scalebox{0.9}[0.9]{$
    \begin{split}
0\le {{P}_{\text{MT},r}}(\Laplace {{t}_{i}})\le P_{\text{MT},r}^{R}-P_{\text{MT},r}^{\text{res}};~
\forall \Laplace {{t}_{i}}\in T_{\text{MT}}^{D}, \forall r\in {{\Omega }_{\text{MT}}}
    \end{split}
    $}
    \label{eqn14}
    \end{equation}
    \begin{equation}
        \scalebox{0.9}[0.9]{$
    \begin{split}
0\le P_{\text{BESS},b}^{C}(\Laplace {{t}_{i}})\le P_{\text{BESS},b}^{\max, C}(\Laplace {{t}_{i}});
    ~\forall \,\Laplace {{t}_{i}}\in T_{\text{BESS}}^{C}, ~\forall \,b\in {{\Omega }_{B}}
    \end{split}
    $}
    \label{eqn15}
    \end{equation}
    \begin{equation}
    \scalebox{0.9}[0.9]{$
    \begin{split}
0 \leq P_{\text{BESS},b}^{D}(\Laplace {{t}_{i}}) \leq  P_{\text{BESS},b}^{\max, D}(\Laplace {{t}_{i}});
    ~\forall \,\Laplace {{t}_{i}}\in T_{\text{BESS}}^{D}, ~\forall \,b\in {{\Omega }_{B}}
    \end{split}
    $}
    \label{eqn16}
    \end{equation}
%    \begin{table*}
\begin{figure*}
	    \begin{equation}
	    \scalebox{0.9}[0.9]{$
    P_{\text{BESS},b}^{C}(\Laplace {{t}_{i}})=
   \begin{cases}
    0;\qquad\text{if}\text{ }{\text{SO{C}}_{b}}(\Laplace {{t}_{i}})={\text{SOC}_{b}^{\max}};\forall \Laplace {{t}_{i}}\in T_{\text{BESS}}^{S},\forall b\in {{\Omega }_{B}} \\ 
    P_{\text{BESS},b}^{\max ,C};\qquad \text{if}~ {\text{SO{C}}_{b}}(\Laplace {{t}_{i-1}})+\frac{{{\eta }_{C}}P_{\text{BESS},b}^{\max ,C}}{W_{\text{BESS},b}^{R}}(\Laplace {{t}_{i}})<{\text{SOC}}_{b}^{\max};\forall \Laplace {{t}_{i}}\in T_{\text{BESS}}^{C},\forall b\in {{\Omega }_{B}} \\ 
    (\text{SOC}_{b}^{\max}-{\text{SO{C}}_{b}}(\Laplace {{t}_{i}}))\frac{W_{\text{BESS},b}^{R}}{\Laplace {{t}_{i}}}; ~~\text{if}~ {\text{SOC}_{b}^{\max}}-{\text{SO{C}}_{b}}(\Laplace {{t}_{i-1}})<\frac{{{\eta }_{C}}P_{\text{BESS},b}^{\max ,C}}{W_{\text{BESS},b}^{R}}\ (\Laplace {{t}_{i}});\forall \Laplace {{t}_{i}}\in T_{\text{BESS}}^{C},\forall b\in {{\Omega }_{B}}
\end{cases}
    $}
    \label{eqn17}
    \end{equation}
%\end{figure*}
%\begin{figure*}
    \begin{equation}
    \scalebox{0.9}[0.9]{$
    P_{\text{BESS},b}^{D}(\Laplace {{t}_{i}})=
    \begin{cases}
    0;\qquad \text{if}~{\text{SO{C}}_{b}}(\Laplace {{t}_{i}})={\text{SOC}_{b}^{\min}};\forall \Laplace {{t}_{i}}\in T_{\text{BESS}}^{S},\forall b\in {{\Omega }_{B}} \\ 
    P_{\text{BESS},b}^{\max ,D};\qquad \text{if}~ {\text{SO{C}}_{b}}(\Laplace {{t}_{i-1}})-\frac{P_{\text{BESS},b}^{\max ,D}}{{{\eta }_{D}}W_{\text{BESS},b}^{R}}(\Laplace {{t}_{i}})\ge{\text{SOC}}_{b}^{\min };\forall \Laplace {{t}_{i}}\in T_{\text{BESS}}^{D},\forall b\in {{\Omega }_{B}} \\ 
    (\text{SO{C}}_{b}(\Laplace {{t}_{i}})-{\text{SOC}_{b}^{\min }})\frac{W_{\text{BESS},b}^{R}}{\Laplace {{t}_{i}}}; \text{ }~~ \text{if}~ {\text{SO{C}}_{b}}(\Laplace {{t}_{i-1}})-{\text{SOC}_{b}^{\min }}<\frac{P_{\text{BESS},b}^{\max ,D}}{{{\eta }_{D}}W_{\text{BESS},b}^{R}}(\Laplace {{t}_{i}});\forall \Laplace {{t}_{i}}\in T_{\text{BESS}}^{D},\forall b\in {{\Omega }_{B}} \\ 
\end{cases}
$}
    \label{eqn18}
    \end{equation}
%        \vspace{-0.8cm}
\end{figure*}
%    \end{table*}
    \begin{equation}
        \scalebox{0.9}[0.9]{$        
    \text{SOC}_{b}^{\min }\le {\text{SO{C}}_{b}}(\Laplace {{t}_{i}})\le \text{SOC}_{b}^{\max }; ~\forall \,\Laplace {{t}_{i}}\in T, ~\forall \,b\in {{\Omega }_{B}}
    $}
    \label{eqn19}
    \end{equation}
    \begin{equation}
        \scalebox{0.9}[0.9]{$    
    \begin{split}
{\text{SO{C}}_{b}}(\Laplace {{t}_{i}})={\text{SO{C}}_{b}}(\Laplace {{t}_{i-1}})+\Bigg( \frac{{{\eta }_{C,b}}P_{\text{BESS},b}^{C}}{W_{\text{BESS},b}^{R}} \\ 
    -\frac{P_{\text{BESS},b}^{D}}{{{\eta }_{D,b}}W_{\text{BESS},b}^{R}}\Bigg)\Laplace t_i; 
        ~~ \forall \,\Laplace {{t}_{i}}\in T_{\text{BESS}}^{C/D}, ~\forall \,b\in {{\Omega }_{B}}
    \end{split}
    $}
    \label{eqn20}
    \end{equation}
    \begin{equation}
        \scalebox{0.9}[0.9]{$
    \text{SOC}_{b}(\Laplace {{t}_{i}})={\text{SO{C}}_{b}}(\Laplace {{t}_{i-1}}); ~\forall \Laplace {{t}_{i}} \in T_{\text{BESS}}^{S},\forall b\in {{\Omega }_{B}}
  $}
    \label{eqn21}
    \end{equation}
    %
%    \begin{table*}
    \begin{equation} %[]
    \scalebox{0.89}[0.9]{$
    {{P}_{\text{rev}}}(\Laplace {{t}_{i}})=
    \begin{cases}
%    \begin{array}{ll}
 0;~\text{If}~{{\delta }_{1}}(\Laplace {{t}_{i}})>\,{{\delta }_{2}}(\Laplace {{t}_{i}});\\
\Re \left( {{V}_{1}}(\Laplace {{t}_{i}})I_{1}^{*}(\Laplace {{t}_{i}}) \right);~\text{otherwise}
%\,{{\delta }_{1}}(\Laplace {{t}_{i}})<\,{{\delta }_{2}}(\Laplace {{t}_{i}});\forall \Laplace {{t}_{i}} %\in {{\Omega }_{s}} \\ 
%    \end{array}
    \end{cases}
  \forall \Laplace {{t}_{i}}\in {T} 
    $}
    \label{eqn22}    
    \end{equation}
%    \end{table*}
 The constraints considered in \eqref{eqn10}--\eqref{eqn13} are representing power flow balance, node voltage limits and branch flow limits of the system. \eqref{eqn14} represents MT unit dispatch constraint. In equation \eqref{eqn10}, `$-$' and `$+$' signs will be used when a BESS is charging and discharging respectively.  Equations \eqref{eqn15}--\eqref{eqn18} represent BESS charging and discharging constraints. BESS SOC limit constraints are presented by \eqref{eqn19}--\eqref{eqn21}. Finally, reverse power flow constraint is defined in \eqref{eqn22}.
 \par A nature-inspired solution methodology employing a recently developed and explored optimization algorithm called as modified African buffalo optimization (MABO) to solve the above optimization problem is briefly discussed in the below section.
%    
%\begin{figure}[h]
%	\includegraphics[width=\columnwidth]{./Pictures/ChromosomeP002.eps}
%	\caption{Generalized structure of an individual adopted in GA}
%	\label{Figure:generalized chromosome}
%\end{figure}
%
%\begin{figure}[h]
%	\centering
%	\includegraphics[width=\linewidth]{./Pictures/FC.eps}
%	\caption{Flow chart of the proposed methodology}
%	\label{Fig:FC}
%\end{figure}
\section{Solution Methodology for Coordinated Scheduling}\label{section:solution methodology}
\begin{figure}
	\centering
	\includegraphics[scale=0.06]{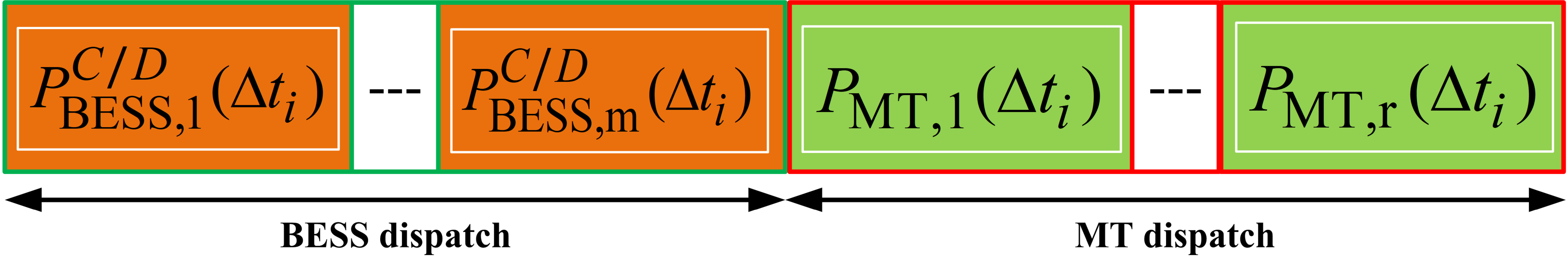}
	\vspace{-0.1cm}
	\caption{Structure of an individual adopted in MABO\vspace{-0.4cm}}
	\label{Figure:generalized chromosome}
\end{figure}

% [nature inspired methods are playing a vital role to solve various real-life problems, which may be very difficult or sometimes impossible to be solved using analytical
%methods. So far, numerous optimization algorithms inspired by
%genetics, nervous systems, and swarm intelligence, based on
%the behavior of birds, fishes, bees, ants, bats, frog, elephant,
%cats, wolf, etc., have been suggested in the literature. These algorithms are applied to solve various complex power system
%optimization problems and are found to be very effective in
%searching the global or near-global solutions. To enhance the
%working of such techniques, various improved or hybrid versions are also available in the literature]
The coordinated scheduling of BESSs and MT unit for maximizing DPF is a complex non-linear multi-constraint
optimization problem which can’t be solved by employing conventional optimization techniques. Such complex combinatorial optimization problems can be solved by using AI based nature inspired meta-heuristic or evolutionary methods which may include genetic algorithm (GA) \cite{meena2017optimal}, gravitational search algorithm (GSA) \cite{singh2019double}, fire works algorithm (FWA) \cite{jadoun2018integration}, African buffalo optimization (ABO) \cite{singh2020modified} and many more. The ABO algorithm was first developed by Odili et al. in 2015 \cite{odili2015african}. It is a swarm intelligence based, nature inspired, and meta-heuristic optimization algorithm motivated by the social and herding behaviour of African buffaloes \cite{wilson1997altruism}. It also offers the capability of solving optimization problems suffering from pre-mature convergence. On the other hand, it suffers from the inability to provide global optimum and to tackle complex engineering optimization problems. To overcome some of these limitations Singh et al. in January of this year \cite{singh2020modified} developed a modified version of ABO named as MABO with the promising potential of offering global or near global optimum solution for complex power system optimization problems besides having well-balanced and well-coordinated exploration and exploitation capabilities. Therefore, in this work MABO is adopted to solve the considered optimization problem. For further details like basic steps, proposed modifications in standard variant, mathematical equations and algorithm parameters’ of MABO the readers may refer \cite{odili2015african,singh2020modified}. The structure of an individual buffalo is composed of decision variables of the problem. The decision variables for the concerned problem include charging/discharging dispatch of BESSs and MT unit dispatch. The generalized structure of an individual for $i$th system state employed in the present work is shown in Fig. \ref{Figure:generalized chromosome}. The figure shows information in terms of dispatches of BESSs and MT units for $i$th system state.

\par
\begin{figure}
	\centering
	\includegraphics[width=\linewidth]{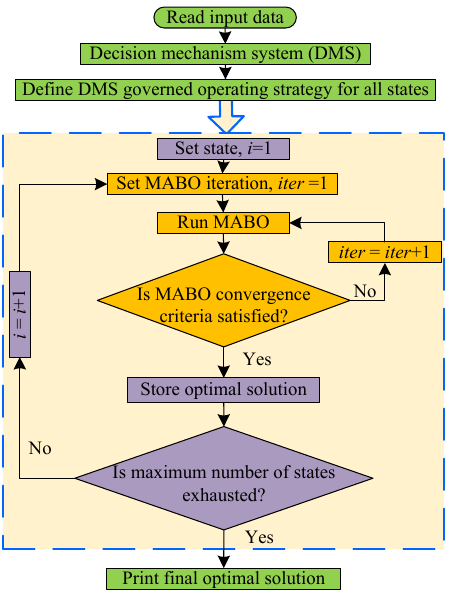}
	\vspace{-0.2cm}
	\caption{Flow chart of the proposed methodology\vspace{-0.6cm}}
	\label{Fig:FC}
\end{figure}

 Decision mechanism system keeps a record of necessary decisions to be taken for system states through mean price based adaptive scheduling, as already discussed in Section \ref{section:decision mechanism}. This provides a feasible environment and guidance for MABO to perform sequential optimization while maximizing DPF of the utility. In addition, it may alleviate the computational burden of MABO by reducing search space. However, the computational efficiency evaluation is beyond the scope of this work. In this way MABO sequentially optimizes all system states in order to evaluate daily profit function of the utility. The flow chart of the proposed methodology, employing MABO as subroutine optimization, is shown in Fig. \ref{Fig:FC}.
\section{Case Study}\label{section:simulation results and discussion}
\subsection{Test system and data}
In this section, the proposed methodology is validated by implementing on 12.66 kV, 33-bus benchmark test distribution system \cite{25627}. For this system, the nominal active and reactive load demand is 3.715 MW and 2.30 MVar, respectively. With this system loading, the power losses and minimum node voltage are 202.67 kW and 0.9131p.u. The bus and line data of this system may be referred from the reference mentioned above. This system is modified by assuming various existing DERs, the sizing and siting of which is taken as shown in Table \ref{table:2}.
 
\begin{figure}
	\centering
	\includegraphics[width=\linewidth]{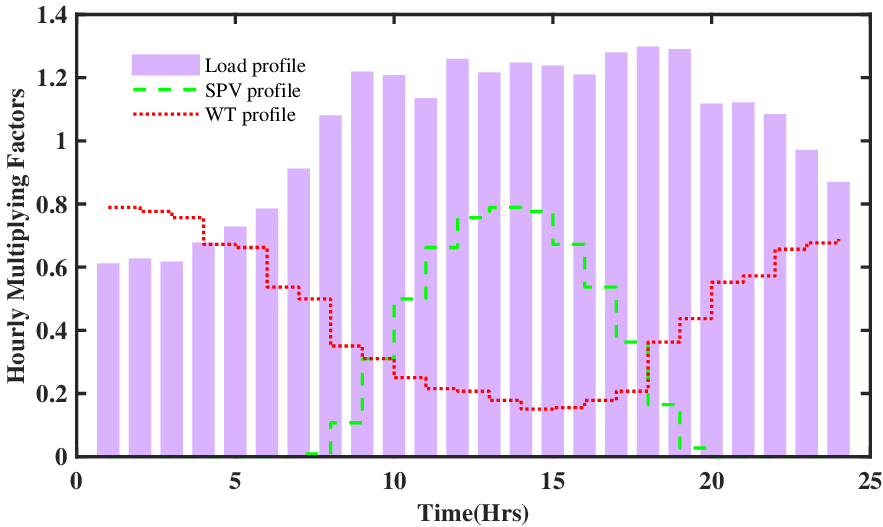}
	\caption{ Synthetic data generated for SPV and WT power generation and load demand}
	\label{Fig:Fig1}
\end{figure}

\begin{figure}
 	\centering
 	\includegraphics[width=\linewidth]{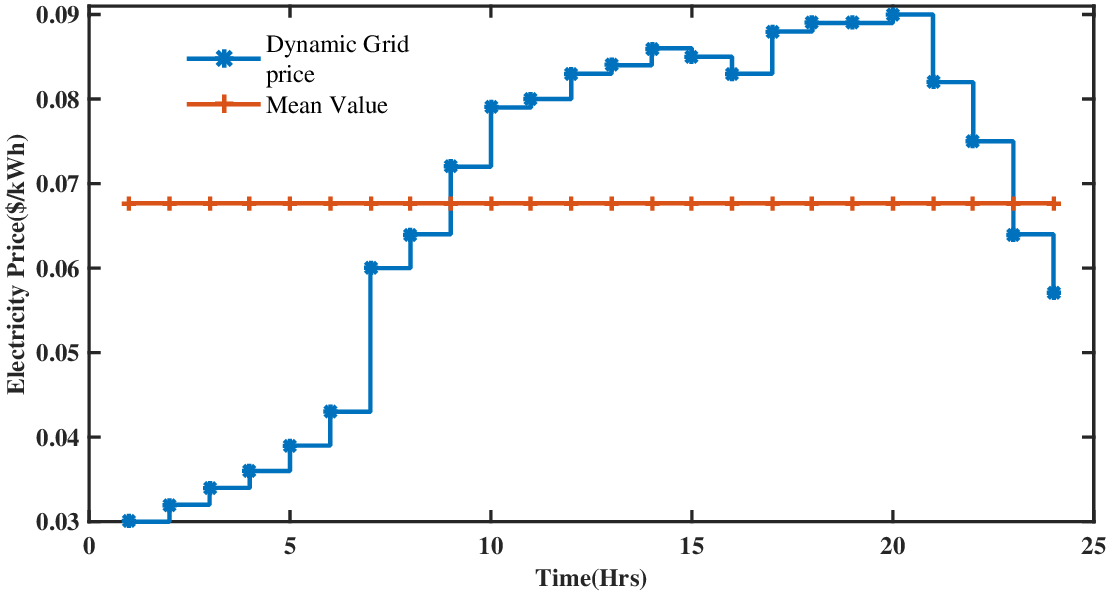}
 	\caption{Day-ahead dynamic grid price over the sample day.}
 	\label{Fig:Fig2}
 \end{figure}

    \begin{table}
  	\centering
  	\caption{Sizing and siting of DERs considered}
  	\label{table:2}
  		\renewcommand{\arraystretch}{1.2}
  	\begin{tabular}{p{2.2cm}M{1.1cm}M{1.1cm}M{1.1cm}}
  		\hline
  		\multirow{2}{*}{\textbf{DER(s)}} & \multicolumn{3}{c}{\textbf{Node(s)}} \\ \cline{2-4} 
  		& 17       & 25           & 30      \\ \hline
  		\textbf{SPV (kWp)}            & 2000    & -       & -       \\ 
  		\textbf{WT (kWp)}             & -       & -       & 2000       \\ 
  		\textbf{MT (kW)}              & -       & 1200       & -       \\ 
  		\textbf{BESS (kWh)}           & 3000    & 3000       & 3000    \\ \hline
  	\end{tabular}
  \vspace{-0.5cm}
  \end{table}
%\begin{table}
%	\centering
%	\caption{Sizing and siting of DERs considered}
%	\label{table:2}
%	\renewcommand{\arraystretch}{1.2}
%	\begin{tabular}{p{2.2cm}M{1.1cm}M{1.1cm}M{1.1cm}M{1.1cm}}
%		\hline
%		\multirow{2}{*}{\textbf{DER(s)}} & \multicolumn{4}{c}{\textbf{Node(s)}} \\ \cline{2-5} 
%		& 17       & 25      & 29      & 30      \\ \hline
%		\textbf{SPV (kWp)}            & 2000    & -       & -       & -       \\ 
%		\textbf{WT (kWp)}             & -       & -       & 2000    & -       \\ 
%		\textbf{MT (kW)}              & -       & 1200    & -       & -       \\ 
%		\textbf{BESS (kWh)}           & 3000    & 3000    & -       & 3000    \\ \hline
%	\end{tabular}
%	\vspace{-0.5cm}
%\end{table}
The reserve on MT unit is taken as 400 kW. The sample day considered for optimal scheduling of BESSs and MT unit is assumed to be composed of 24 states each of duration one hour. For SPV and WT, the power generation data on hourly basis is taken from \cite{nrel} while as \cite{posoco} may be referred for hourly data of load consumption. The available data is normalized with respect to the rated capacity of considered DERs and peak load demand (taken as 1.3 times the nominal system demand) so as to get hourly generation and load multiplying factors. The dynamic load multiplying factors are assumed to be similar among all buses for a particular system state. These load and renewable generation multiplying factors are further utilized for synthetic data generation by using modified uncertainty sets already discussed in Section \ref{section:RELC Modeling}. The profiles of synthetic data generated for load demand and SPV and WT power generation are shown in Fig. \ref{Fig:Fig1}.
 A dynamic energy pricing scheme is considered, to be offered by the transmission network operators. The sale price of energy to customers is also considered dynamic but is assumed 5$\%$ higher for each system state. The day-ahead time varying price of energy purchase from the grid is taken from \cite{IEX}, as shown in Fig. \ref{Fig:Fig2}. The figure also shows mean price line thus defines the sub-period $T_{\text{BESS}}^{D}$ from 09:00 Hrs to 23:00 Hrs and $T_{\text{BESS}}^{C}$  for the remaining hours of the day.
 Various other parameters considered for simulation purpose are presented in Table \ref{Table:4}. The node voltage constraint limits are taken as $\pm$5$\%$. Simulations are executed on a personal computer of Intel\textsuperscript{\textregistered} Core i5-6600 CPU, 3.30GHz processor with random access memory of 4GB. The results obtained for the optimal scheduling of BESSs and MT unit are presented and investigated.
  \begin{table}[!ht]
  	\centering
  	\caption{Simulation parameters used in the proposed study}
  	\label{Table:4}
  		\renewcommand{\arraystretch}{1.2}
  	\scalebox{0.95}[0.95]{
  	\begin{tabular}{p{1.5cm}M{2.8cm}M{1.5cm}M{1.8cm}}
  		\hline
  		\textbf{Simulation Parameter}      & \textbf{Value}                       & \textbf{Simulation Parameter} & \textbf{Value}                         \\ \hline
  		$H$                                          & $\rm 85\%$                           & $\rm M_{\text{BESS}}$                        & 0.00150  \$/kWh \\ 
  		$\rm P_{\text{BESS,b}}^{\max,C},\ P_{\text{BESS,b}}^{\max,D} $ & 500kW, 500kW                         & $\rm M_{\text{MT}}$                          & 0.012 \$/kWh   \\ 
  		$\rm {\text{SOC}_b^{\min}}, \ {\text{SOC}_b^{\max}}$           & 0.1, 1.0                             & $\rm E_{\text{SPV}}$                         & 0.028 \$/kWh   \\ 
  		$\rm {\text{SOC}_{I}}$                              & 0.1                                  & $\rm E_{\text{WT}}$                          & 0.029 \$/kWh   \\ 
  		$\rm E_{\text{BESS}}^{F}$                         & 0.067 \$/kWh & $\rm E_{\text{MT}}$                          & 0.0335  \$/kWh  \\ 
  		\hline
  		\multicolumn{4}{l}{\tiny where, $\eta_{C} = \eta_{D} = \sqrt{H}$}
  	\end{tabular}
}
\vspace{-0.7cm}
  \end{table}
 \subsection{Simulation results, validations and discussions}
 In this section, the proposed optimization problem is solved by using MABO presented in Section \ref{section:solution methodology}. The pertinent parameters of MABO are taken as: the population size is 10, and the maximum number of generations is 100. A Backward/Forward sweep method is employed to solve the load flow equations. The optimal economic equation obtained under proposed framework (named as `A') is presented in Table \ref{Table:5}. It shows that total payment incurred in purchasing energy from the grid and renewable DGs, and in producing energy from the MT unit and BESSs is US$\$$ 5395.14$\approx$5.40k. The revenue generated by billing of customers, including the cost of feeder power loss, is US$\$$ 7593.42$\approx$7.60k. The profit function of the sample day considered is therefore US$\$$ 2198.28$\approx$2.20k which is around 41$\%$. However, the actual profit will be less after deducting the returns against the installation cost of BESSs and MT unit, taxes, etc. The table also shows that almost three fourth of the payment is incurred for grid power purchase and the rest one fourth dispersed among local generations.
    \par The energy equation obtained for proposed optimal scheduling of BESSs and MT is also presented in Table \ref{Table:6}. The table shows that the system deals with 110587.59 units (kWh) of energy out of which about 56$\%$ is imported from the grid and 44$\%$ is matched locally. The contribution of SPV and WT units is about 10$\%$ and 20$\%$, respectively, whereas BESSs and MT unit each contributed about 7$\%$. The self-adequacy of the system is of the order of 44$\%$. The table also shows that charging of BESSs requires 10364 kWh but delivers 7450 kWh in the system. For the given study, the price of unit purchase from the local resources is taken cheaper than the grid dynamic price. Therefore, about 75$\%$ of the total payment needed to match grid purchase which is only 56$\%$ of the total energy demand.
   \par
\begin{figure*}
	\centering
	\subfloat[BESS at node 17\label{Fig:4a}]{\includegraphics[width=0.33\linewidth]{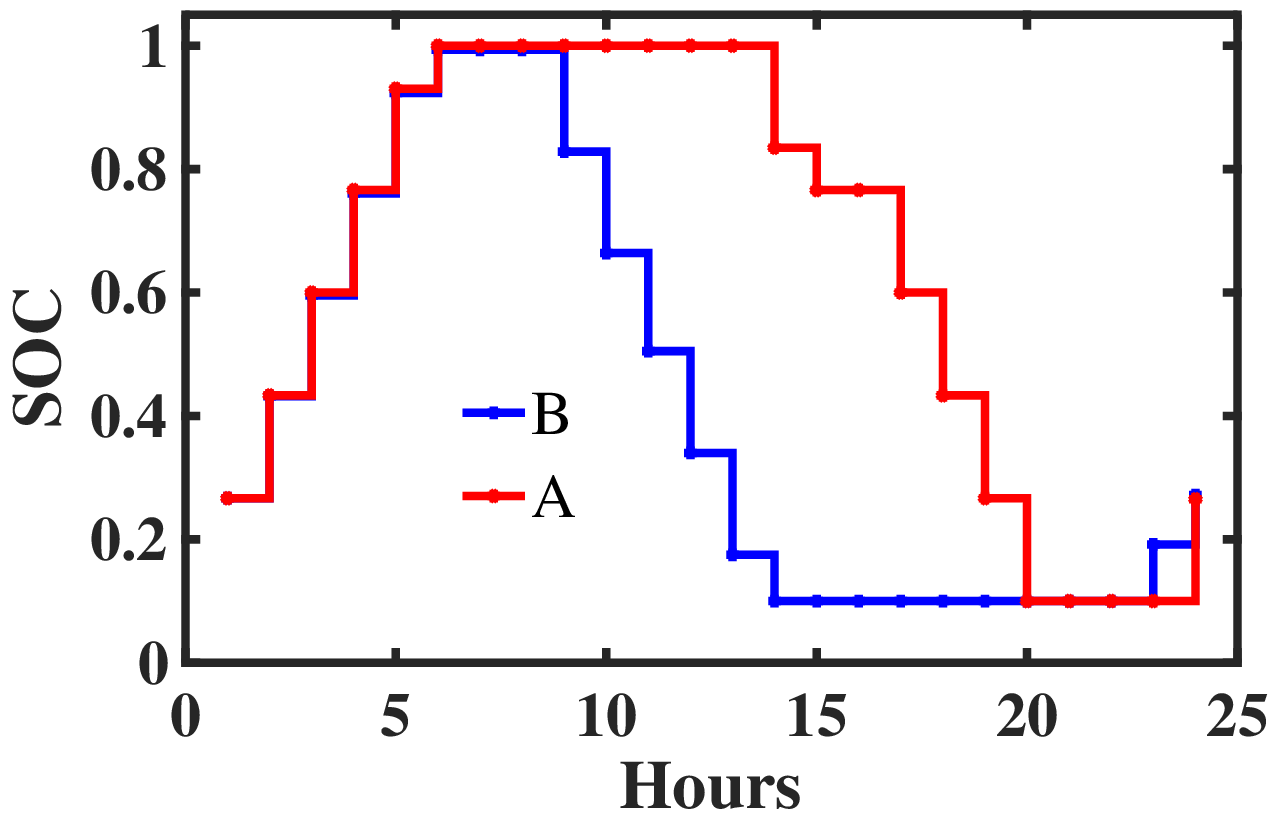}}	
	\subfloat[BESS at node 25\label{Fig:4b}]{\includegraphics[width=0.33\linewidth]{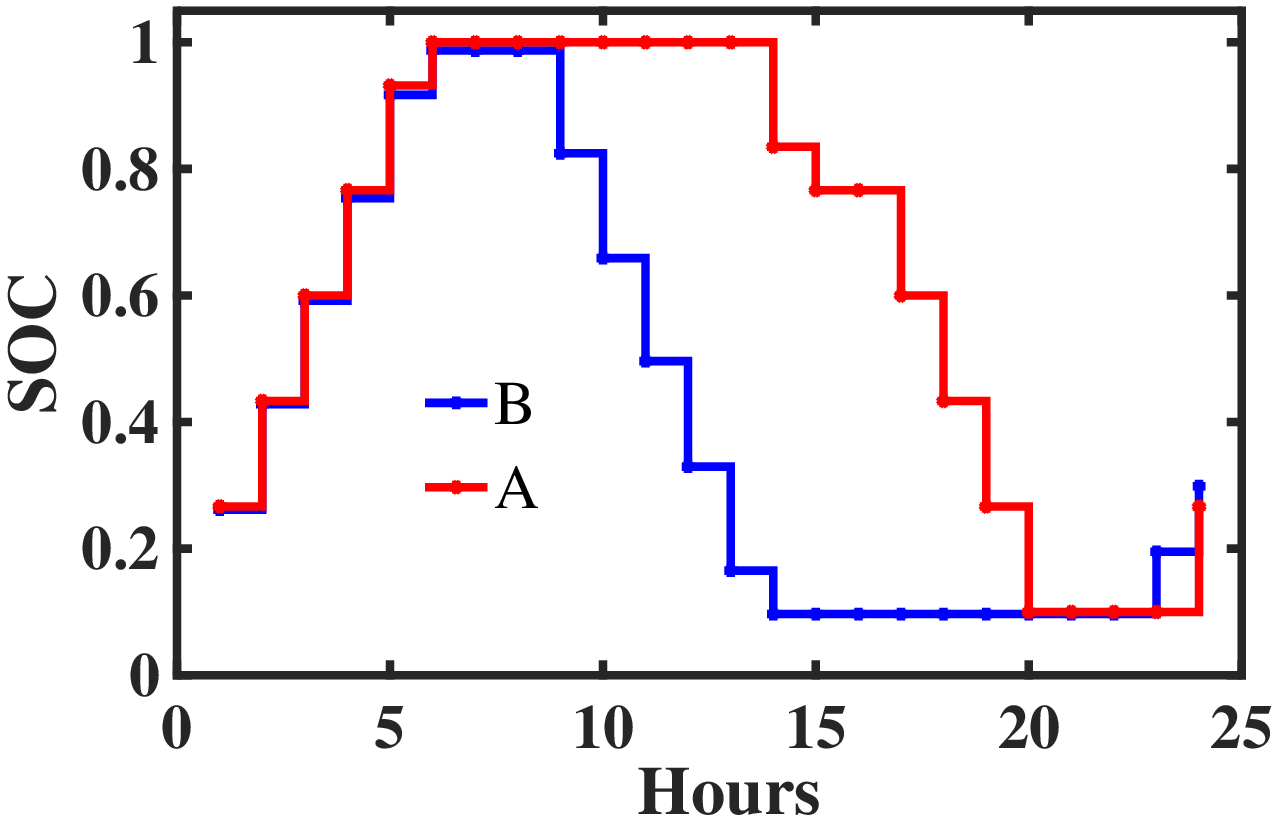}}		
	\subfloat[BESS at node 30\label{Fig:4c}]{\includegraphics[width=0.33\linewidth]{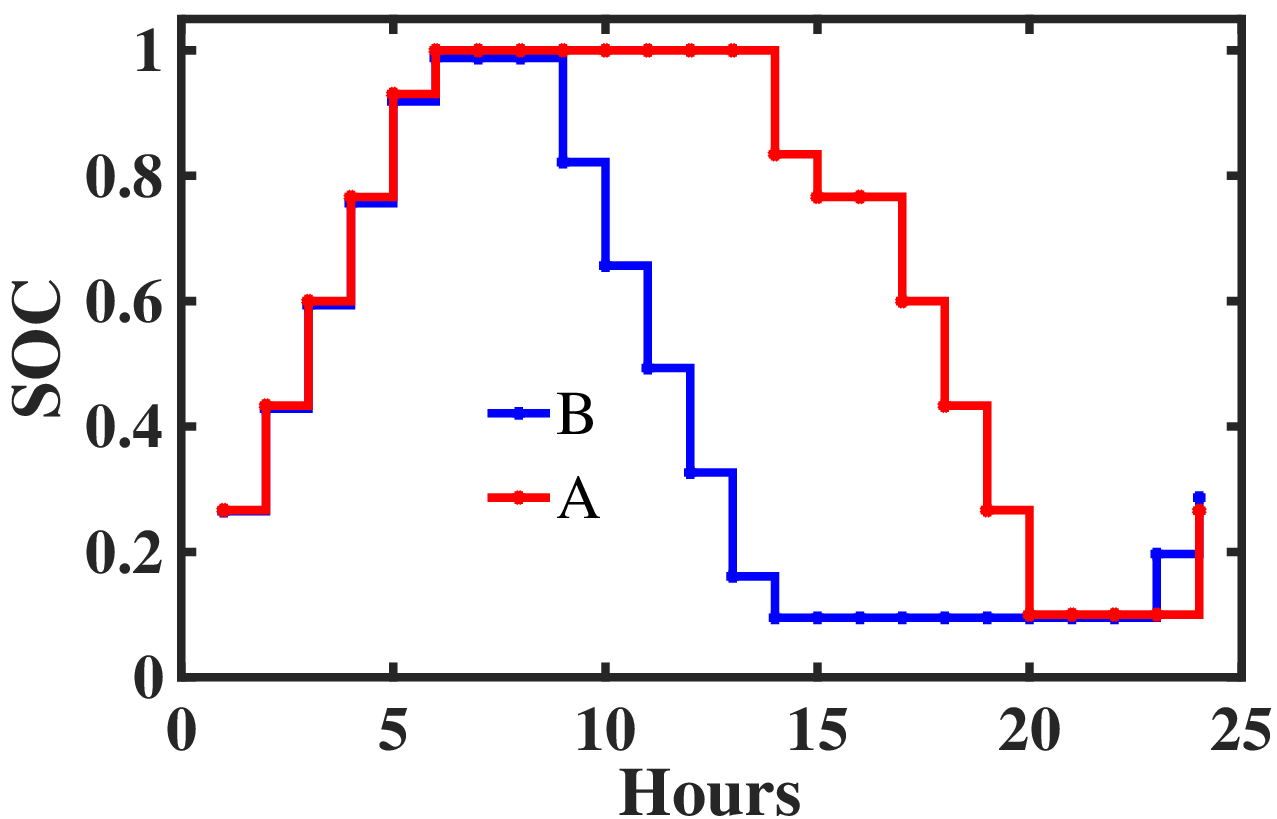}}
	\vspace{-0.2cm}
	%\subfloat[]{\includegraphics[width=0.5\linewidth]{./Pictures/Fig4d.eps}}	
	\caption{Optimal dispatch/SOC of BESSs obtained by proposed approach (A) and existing approach (B)\cite{5438853}.	\vspace{-0.5cm}}
	\label{Fig:4}
\end{figure*}

 To validate the promising profit maximization abilities of proposed optimization framework, the obtained simulation results are compared with a well-established BESS operation strategy suggested in \cite{5438853}, named as `B'. For this purpose the strategy of \cite{5438853} is applied to this system and the results obtained are compared with the proposed strategy as presented in Table \ref{Table:5} and Table \ref{Table:6}. It has been observed that the daily profit is 3.94$\%$ more and the losses are 1.50$\%$ less using the proposed strategy. The charging/discharging of BESSs using these two strategies is compared in Fig. \ref{Fig:4}. It can be observed from the figure that the charging of BESSs is almost same using both the strategies; however, the discharging is quite different. The strategy of \cite{5438853} discharges BESSs immediately after full charging that shifts standby mode of BESSs to peak hours. Thus this strategy is unable to exploit peak shaving and associated technical benefits. But, proposed strategy suitably manages standby periods of BESSs in such a way that they essentially discharge during on-peak hours. This shows that BESSs can be managed in a better way using proposed MPAS.
   \vspace{-0.5cm}
    \begin{table}[!ht]
  	\centering
  	\caption{Comparison of economic equation over the scheduling period (in US\$$\times 10^3$ and \%) \label{Table:5}}
  	  		\renewcommand{\arraystretch}{1.2}
  	\scalebox{0.9}[0.9]{
  		\begin{tabular}{|p{0.07cm}|P{0.45cm}|P{0.4cm}|P{0.4cm}|P{0.35cm}|P{0.45cm}|P{0.5cm}|P{0.3cm}|P{0.5cm}|P{0.38cm}|P{0.3cm}|P{0.45cm}|}
  			\hline
  			\multirow{2}{*}{\textbf{\rotatebox[]{90}{\parbox{1.6cm}{Method}}}}& \multicolumn{7}{|c|}{\textbf{Payments}} & \multicolumn{3}{c|}{\textbf{Revenue}}& \multirow{2}{*}{\rotatebox[]{90}{\textbf{\parbox{1.6cm}{Profit}}}} \\
  			\cline{2-11}
  			& \textbf{Grid} & \textbf{SPV} & \textbf{WT} & \textbf{MT}  & \textbf{MT (OM)} & \textbf{BESS (OM)} & \rotatebox[]{90}{\textbf{Total}} & \textbf{\rotatebox[]{90}{\parbox{1.6cm}{Cost of \\consumption}}} & \textbf{\rotatebox[]{90}{\parbox{1.2cm}{Cost of\\ losses}}}  & \rotatebox[]{90}{\textbf{Total}} & \\
  			\hline
  			A & 4.06/ 75.21 & 0.32/ 5.89 &0.63/ 11.66 & 0.27/ 4.97 & 0.10/ 1.78 & 0.03/ 0.49  & 5.40& 7.34/ 96.69& 0.25/ 3.31& 7.60& 2.20/ 40.75\\
  			\hline
  			B & 4.15/ 75.63 & 0.32/ 5.78 & 0.63/ 11.45 & 0.27/ 4.88 & 0.10/ 1.74 & 0.03/ 0.48  & 5.49& 7.34/ 96.58 & 0.26/ 3.41& 7.60 & 2.11/ 38.46\\
  			\hline
  			\multicolumn{11}{l}{\tiny A$\rightarrow$ Proposed method, \qquad B$\rightarrow$ Method of \cite{5438853} }
  		\end{tabular}
  	}
  \end{table}
  \begin{table}[!ht]
  	\centering
  	\caption{Comparison of energy equation over the scheduling period (in MWh and \%)}
  	\label{Table:6}
  	\renewcommand{\arraystretch}{1.2}
  	\scalebox{0.9}[0.9]{
  		\begin{tabular}{|p{0.07cm}|p{0.52cm}|P{0.52cm}|P{0.52cm}|P{0.4cm}|P{0.5cm}|P{0.6cm}|P{0.58cm}|P{0.4cm}|P{0.5cm}|P{0.6cm}|}
  			\hline
  			\multirow{2}{*}{\rotatebox[]{90}{\parbox{1.0cm}{\textbf{Method}}}}&\multicolumn{3}{c|}{\textbf{Purchased}} & \multicolumn{2}{c|}{\textbf{Supplied}}& \multirow{3}{*}{\textbf{Total}}&\multicolumn{3}{c|}{\textbf{Consumed}} & \multirow{3}{*}{\textbf{Total}} \\
  			\cline{2-6} 
  			\cline{8-10}
  			&\multirow{2}{*}{\textbf{Grid}}& \multirow{2}{*}{\textbf{SPV}}& \multirow{2}{*}{\textbf{WT}}& \multirow{2}{*}{\textbf{MT}} & \multirow{2}{*}{\textbf{BESS}} & &\multirow{2}{*}{\textbf{Billing}}& \multirow{2}{*}{\textbf{Loss}}& \multirow{2}{*}{\textbf{BESS}}& \\
  			&&&&&&&&&&\\
  			\hline
  			A & 62.10/ 56.15 & 11.35/ 10.26 & 21.69/ 19.61 & 8.00/ 7.23   & 7.45/ 6.74  & 110.59  & 96.68/ 87.42  & 3.55/ 3.20& 10.36/ 9.37 & 110.59\\
  			\hline
  			B & 62.33/ 56.28 & 11.35/ 10.25 & 21.69/ 19.58 & 8.00/ 7.22 & 7.36/ 6.65 & 110.73 & 96.68/ 87.30 &3.60/ 3.25 &10.45/ 9.43 & 110.73\\
  			\hline
  			\multicolumn{11}{l}{\tiny A$\rightarrow$ Proposed method, \qquad B$\rightarrow$ Method of \cite{5438853}} 
  		\end{tabular}
  	}
  \end{table}
%\vspace{-0.5cm}
%
% \begin{figure}
%	\centering
%	\includegraphics[width=\linewidth]{./Pictures/Fig3.eps}
%	\caption{Profit function for system states using optimal solution.}
%	\label{Fig:Fig3}
%\end{figure}
\par
\begin{figure}[!ht]
	\centering
\includegraphics[width=\linewidth]{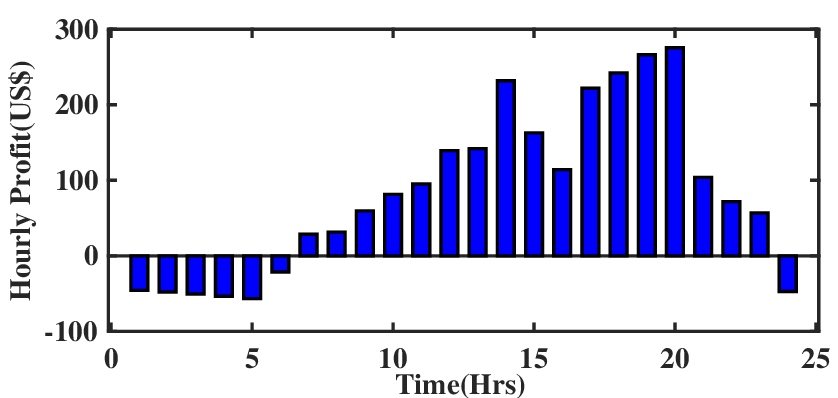}
\caption{Values of hourly profit function obtained under proposed optimization framework \label{Fig:5}}
\end{figure}

 After validation of the proposed optimization framework, its promising features has been investigated further by presenting the analysis of each state. The optimal solution provides profit function for each state as presented in Fig. \ref{Fig:5}. It can be observed that the profit function becomes negative during certain off-peak hours. It happened because additional energy is being drawn from the grid or DGs against the charging of BESSs thus increases payments and supersedes revenue owing to lesser off-peak demand. The DPF, however, is found to be US$\$$ 2198.28 with positive sign, as also shown in Table \ref{Table:5}. A dip in the profit function is observed during sub-period $T_{\text{BESS}}^{D}$ which is on account of dip in load demand as well as dynamic price and standby mode of all BESSs, otherwise it maintains fairly higher positive values during most of the peak pricing hours, i.e. 14:00 Hrs and 17:00 Hrs. to 20:00 Hrs. 
\par
\begin{figure}[!ht]
	\centering
	\includegraphics[width=\linewidth]{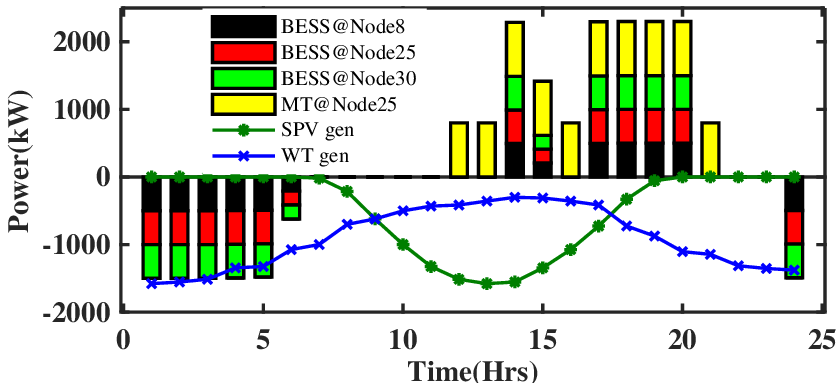}
	\caption{Optimal scheduling of all BESSs and MT\label{Fig:6}}
\end{figure}

 The optimal scheduling and dispatches of BESSs and MT unit obtained using proposed method is presented in Fig. \ref{Fig:6}. It can be observed that all BESSs exclusively charges and discharges during sub-periods $T_{\text{BESS}}^{C}$ and $T_{\text{BESS}}^{D}$, respectively and these two sub-periods are separated by standby period.  As shown in Fig. \ref{Fig:4}, all BESSs are fully utilized as their SOC status reached to the same level in one round trip. Total energy dispatch from these DERs is found to be about 32$\%$ of the energy demand that prevails during on-peak hours, i.e. 12:00 Hrs. to 21:00 Hrs. It can be observed that BESSs charge mostly from WT and discharges during peak pricing hours where solar power is not much available. Thus BESSs neutralize the intermittency of wind power and delivers the same during acute operating conditions of the system, besides gaining charge differential benefits. The power generated from SPV and WT are deliberately shown with negative sign merely to have better understanding.
%
 %
%  \begin{figure}[!ht]
% 	\centering
% 	\includegraphics[width=\linewidth]{./Pictures/Fig5.eps}
% 	\caption{Effect of optimal scheduling of BESSs and MT on load profile.}
% 	\label{Fig:Fig5}
% \end{figure}   
    %
   \begin{figure}[!ht]
  	\centering
  	\includegraphics[width=\linewidth]{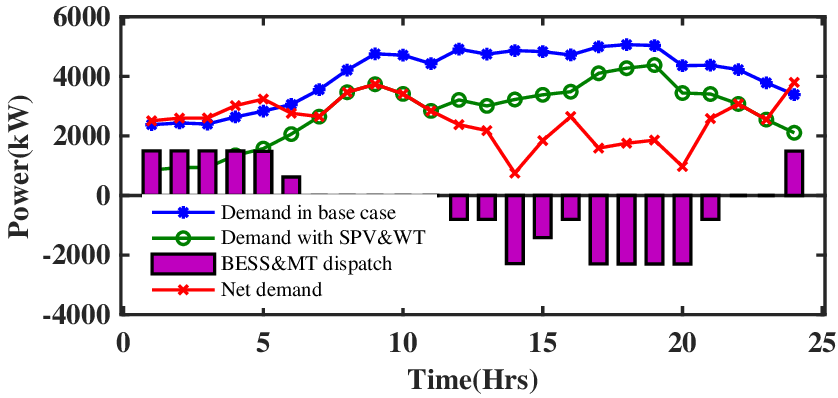}
  	\vspace{-0.2cm}
  	\caption{Load profile of the system under different scenarios\label{Fig:7}} 
  	\vspace{-0.6cm}
  \end{figure} 
 \par The modification in load profile while employing optimal dispatches from BESSs and MT is shown in Fig. \ref{Fig:7}. The figure shows load profile without and with renewable DGs, and also the net load profile while considering optimal dispatches from BESSs and MT unit. It can be observed that the fluctuations in load profile increases using renewables and that further worsen using BESSs and MT unit. However, mean and peak demand are found to be reduced by 9.87$\%$ and 14.71$\%$ whereas the valley demand is reduced by 11.99$\%$ using optimal scheduling of BESSs and MT. Moreover, the peak demand is shifted from 19:00 Hrs. to 09:00 Hrs., whereas the new valley demand occurred at 14:00 Hrs. instead of 01:00 Hrs. Another useful finding is observed while comparing the load profiles. The load deviation index (LDI) is evaluated for these profiles. The index is standard deviation of the demands constituting the load profile. Smaller value of this index is desirable. The LDI of load demand is found to be 924.38 kW which is deteriorated to 1031.07 kW by the integration of SPV and WT units. However, it is improved to 765.45 kW using optimal scheduling of BESSs and MT unit though the scheduling was determined on economic basis. Thus pure economic optimal scheduling of BESSs and MT causes moderate peak demand shaving and valley deepening, partial rebound effect and enhanced LDI, but with somewhat fluctuating load demand. The fluctuations in load demand attributed to maximizing price differential benefits using proposed scheduling during peak pricing hours. The net demand depict no reverse power flow thus ensures better coordination of existing energy resources.
    \section{Conclusions}
    \label{section:conclusions}
    The paper addresses optimal utilization of existing dispatchable and non-dispatchable energy resources by coordinating utility owned BESSs and MT unit to maximize daily profit function (DPF) while accounting for uncertainty in non-dispatchable resources (SPV \& WT) and load demand. Proposed methodology introduces mean price-based adaptive scheduling (MPAS) being embedded within a decision mechanism system (DMS). Proposed algorithm and fictitious charges for BESS operation in dynamic pricing provides necessary information to DMS. DMS process the information as a priori for possible maximization of profit function thus guides MABO for the concurrent state. The application results obtained on a standard test bench shows that MPAS provides economic system operation, better exploitation of available energy resources and improves LDI. The comparison of the proposed strategy with the existing one shows enhanced profit of the utility and better utilization and management of existing energy resources besides power loss reduction. However, peak hours may face some fluctuations in load demand on account of gaining price-differential benefits. This highlights possible limitation that may arise while employing pure economic optimal scheduling of BESSs and MT. Either multi-objective optimization or ancillary services may overcome this limitation. Demand response is ignored in the present work but can play vital role in deciding load profile and load fluctuations. The present work may be extended to explore these concerns. Further, it can be concluded that the proposed flexible and adaptive scheduling strategy will be promising for operating BESSs in the future energy networks.
	\balance 
	\bibliographystyle{IEEEtran}
	\bibliography{IEEEreferences}
\end{document}